\documentstyle[12pt]{article}
\begin{document}
\vspace*{120pt}
\noindent
{\bf NONISOSPECTRAL SYMMETRIES OF THE KDV EQUATION}
\vspace{12pt}
\par
\noindent
{\bf AND THE CORRESPONDING SYMMETRIES}
\vspace{12pt}
\par
\noindent
{\bf  OF THE WHITHAM EQUATIONS. }\footnote{This text wes published in
the book ``Singular Limits of Dispersive Wawes'' eds. N.M.Ercolany,
I.R.Gabitov, C.D.Levermore, D.Serre, Plenum Press, NY, 1994.}
\vspace{36.0pt}
\par
\noindent
\hspace{1.0truein}P.G.Grinevich
\vspace{12.0pt}
\par
\noindent
\hspace{1.0truein}L.D.Landau Institute for Theoretical Physics,
\par
\noindent
\hspace{1.0truein}Kosygina 2, Moscow, 117940, GSP-1, USSR.
\vspace{36.0pt}
\par
\noindent
{\bf 0. INTRODUCTION.}
\vspace{12.0pt}
\par
In our paper we construct a new infinite family of symmetries of the
Whitham equations (averaged Korteveg-de-Vries equation). In contrast
with the ordinary hyd\-ro\-dy\-na\-mic-type flows these symmetries are
nonhomogeneous (i.e. they act nontrivially at the constant solutions),
are nonlocal, explicitly depend upon space and time coordinates and
form a noncommutative algebra, isomorphic to the algebra of the
polynomial vector fields in the complex plane.
\par
We will study the averaged Korteveg-de-Vries (KdV) equation.
But the main results of our paper can be easily extended to other
averaged systems, associated with the integrable by the inverse
scattering transform $1+1$ equations.
\par
We will assume original KdV to be normalized as:
$$
u_t= \frac {1}{4} u_{xxx}-\frac{3}{2} uu_x.
\eqno (0.1)
$$
We will also need a $2+1$ - dimensional generalization of KdV - the
Kadomtsev - Petviashvili (KP) equation
$$
(u_t - \frac {1}{4} u_{xxx}+\frac{3}{2} uu_x)_x = 3 {\alpha }^2 u_{yy},\
{\alpha }^2 = \pm 1.
\eqno (0.2)
$$
\par
Let us recall the definition of the averaged KdV equations in the classical
one-phase case (see [1] for more detailes). Equation (0.1) possesses a
three-parametric family of moving waves
$$
u(x,t) = 2 {\wp} (x - vt|g_2,g_3)- v/6.
\eqno (0.3)
$$
(${\wp} (x,|g_2,g_3)$ denotes the Weierstrass ${\wp}$ - function).
We may try to construct asymptotical KdV solutions of the modulated
wave type
$$
u(x,t) = 2 {\wp} (x - v(X,T)t+\phi(X,T)|g_2(X,T),g_3(X,T))- v(X,T)/6+\epsilon
u_1,\
\eqno (0.4)
$$
where $X=\epsilon x$, $T=\epsilon t$,
$\epsilon$ is a small parameter ($X$ and $T$ are called "slow" variables).
If the correcting term $u_1$ is a bounded function as
$x \sim 1/\epsilon, t \sim 1/\epsilon$ then it can be proved (see [1], [3])
that
the functions $g_2(X,T)$, $g_3(X,T)$, $v(X,T)$ satisfy some
hydrodynamical-type equations:
$$
\frac{\partial}{\partial T} \left ( \begin{array}{c} g_2(X,T) \\ g_3(X,T) \\
v(X,T) \end{array} \right ) = {\hat V} (g_2,g_3,v) \cdot
\frac{\partial}{\partial X} \left ( \begin{array}{c} g_2(X,T) \\ g_3(X,T) \\
v(X,T) \end{array} \right )
\eqno (0.5)
$$
where the matrix ${\hat V} (g_2,g_3,v)$ can be expressed via elliptic
functions.
\par
The averaged KdV (0.5) may be transformed to the Riemann diagonal form
$$
\frac{\partial}{\partial T} \left ( \begin{array}{c} r_1(X,T) \\ r_2(X,T) \\
r_3(X,T) \end{array} \right ) =
\left ( \begin{array}{ccc} v_1(\vec r) & 0 & 0 \\ 0 & v_2(\vec r) & 0 \\
0 & 0 & v_3(\vec r) \end{array} \right )
\frac{\partial}{\partial X} \left ( \begin{array}{c} r_1(X,T) \\ r_2(X,T) \\
r_3(X,T) \end{array} \right ).
\eqno (0.6)
$$
The diagonalizing change of variables (Whitham [1]) is the following :
the Riemann invariants $r_i$, $i=1,2,3$ are the roots of the polynomial
$R(\lambda)= 4 (\lambda + v/6)^3-g_2 (\lambda + v/6) - g_3$,
$r_1 \leq r_2 \leq r_3$. The  matrix ${\hat V}$ in the diagonal form reads as:
$$
v_1(\vec r)=\frac{r_1+r_2+r_3}{3}-
\frac{2}{3} (r_2-r_1) \frac {K(s)}{K(s)-E(s)}
$$
$$
v_2(\vec r)=\frac{r_1+r_2+r_3}{3}-
\frac{2}{3} (r_2-r_1) \frac {(1-s^2)K(s)}{E(s)-(1-s^2)K(s)}
\eqno (0.7)
$$
$$
v_3(\vec r)=\frac{r_1+r_2+r_3}{3}+
\frac{2}{3} (r_3-r_1) \frac {(1-s^2)K(s)}{E(s)}
$$
where $s^2=(r_2-r_1)/(r_3-r_1)$, $E(s), \ K(s)$ are elliptic integrals.
\par
Averaging procedure can be treated as a nonlinear analog of the WKB method.
It can also be applied to the multiphase solutions of soliton equations
(nonlinear superpositions of moving waves) (Flaschka - Forest - McLaughlin
[2], Lax - Levermore [3]). Averaged KdV equations are known as Whitham
equations. Averaged equations associated with soliton systems are also known
as equations of slow modulations or as equations of the soliton lattice
dynamics.
\par
It is well-known that the original KdV equation is an integrable in the
Liouville sense hamiltonian system. For averaging (at least in the
one-phase case) the integrability is not necessary but additional structures
of the original equations are usually inherited in the associated averaged
equations. (The connection between the original structures and the averaged
ones may be very nontrivial). The Whitham equations, associated with the KdV
solutions with arbitrary number of phases have the following properties.
\par
1) They can be presented in the hamiltonian form (Dubrovin - Novikov [4]).
\par
2) They can be written in the Riemann diagonal form (Flaschka - Forest -
McLaugh\-lin [2]). For the hydrodynamical-type systems with more then
2 components the existence of the Riemann invariants is a very nontrivial
property.
\par
3) They have an infinite number of conservation laws in involution
(i.e. an infinite number of mutually commuting symmetries). The full
set of integrals consist of $2g+1$ infinite families where $g$ is the
number of phases. One of these families is formed by the averaged
higher KdV flows, the other $2g$ have no direct analogs in the KdV theory.
These $2g$ additional families were constructed by S.P.Tsarev [5] for $g=1$ and
by B.A.Dubrovin [6] for all $g$.
\par
4) They can be integrated by the generalized hodograph transform (Tsarev [7],
[5]). In fact Tsarev proved that any hamiltonian hydrodynamical-type $1+1$
system possessing Riemann invariants is integrable. Algebro-geometrical
interpretation of the hodograph transform was suggested by I.M.Krichever [8].
\par
The fact that the KdV equation has an infinite set of commuting symmetries
(higher KdV) is well-known. They can be expressed via recursion operator
$\Lambda$
$$
\frac {\partial u}{\partial t_{2n+1}}=\frac {d}{dx}
{\left(-\frac {\Lambda}{4}\right)}^n
u = -2\frac{d}{dx}{\left(-\frac {\Lambda}{4}\right)}^{n+1} \cdot 1.
\eqno (0.8)
$$
where $\Lambda=-\partial _x^2+2\partial_x^{-1} u \partial_x +2u$,
$x=t_1$, $t=t_3$. But it is less known that the group of symmetries
of KdV is much wider (the same is valid for all integrable by the inverse
scattering transform systems). In our parer the averaged KdV symmetries,
associated with the following KdV symmetries will be studied
$$
\frac {\partial u}{\partial \beta_{2n}}=-2\frac {d}{dx}
{\left(-\frac{\Lambda}{4}\right)}
^{n+1}(\sum_{k=0}^{\infty}((2k+1)t_{2k+1}
{\left(-\frac {\Lambda}{4}\right)}^k)\cdot 1.
\eqno (0.9)
$$
\par
Let all the the higher times $t_k$, $k>3$ be equal to zero.
Then for $n=-2,0,2$ we have:
$$
\partial u / \partial \beta _{-2}= 3tu_{x} -2\qquad \hbox {(Galilean).}
\eqno (0.10)
$$
$$
\partial u / \partial \beta _{0} =
3 t (\frac {1}{4} u_{xxx} - \frac {3}{2} u u_x ) +  x u_{x} + 2 u
\qquad \hbox {(Scaling).}
\eqno (0.11)
$$
$$
\partial u / \partial \beta _{2} =
3 t (\frac {1}{16} u^{\rm v} - \frac {5}{8} u u_{xxx} - \frac {5}{4} u_x u_{xx}
+ \frac {15}{8} u^2 u_x) +
x (\frac {1}{4} u_{xxx} - \frac {3}{2} u u_x ) +
u_{xx} - 2 u^{2} - \frac {1} {2} u_{x} \partial^{-1}_x u.
\eqno (0.12)
$$
\par
We call the symmetries (0.9) nonisospectral because they change the spectrum
of the auxiliary linear problem (see section 2) or Virasoro because they
form a noncommutative algebra, isomorphic to the algebra of the polynomial
vector fields in the complex plane (the full Virasoro algebra does emerge
in the KP theory). The ordinary higher KdV are isospectral in this sense.
\par
First examples of the nonisospectral integrable equations (without connection
with the symmetry problem) were constructed by F.Calogero - A.Degasperis [9],
V.A.Belinskii - V.E.Zakharov [10], Maison [11], G.Barucchy - T.Regge [12].
The fact that we can generate new KdV symmetries by applying the recursion
operator to the scaling transformation was pointed out by
N.H.Ibragimov - A.B.Shabat in [13], but they did not pay much attention
to this fact because they studied in [13] only local symmetries. Some
nonisospectral KP symmetries were constructed by H.H.Chen, Y.C.Lee, J.E.Lin
[14], F.J.Schwarz [15]. A general approach based on the so-called
mastersymmetries was developed by B.Fuchssteiner and W.Oevel [16]. A very
convenient method of constructing nonisospectral symmetries for the integrable
by the inverse scattering transform equations was developed by A.Yu.Orlov and
E.I.Schulman [17]. An analog of formula (0.9) for the KP equation was found
by A.S.Fokas and P.M.Santini [18].
\par
All these papers were dedicated to the algebraic theory of the nonisospectral
symmetries or to the vanishing in the infinity case. The periodic (finite-gap)
theory needs a special consideration. I.M.Krichever and S.P.Novikov calculated
the action of a part of such symmetries on the finite-gap KP solutions in [19]
(their subalgebra was associated with a given KP solution and did not vary
the spectral curve). They posed also the problem for the full KP analog of the
algebra (0.9).
\par
The action of the symmetries (0.9) on the finite-gap KdV solutions was
calculated by A.Yu.Orlov and the author in [20], [21]. Similar results were
also obtained in [20], [21] for the KP equation. It was shown that this
action has a natural geometrical interpretation (see paragraph 1.3).
\par
Late 80'ies a number of links between the nonlinear integerable equations
and the conformal field theory was discovered. For example the finite-gap
$\tau$-function in the KP theory coincides with the determinant of the
$\bar \partial$ operator on the appropriate bundle (see, for example [20]).
This determinant plays the central role in the string theory. The matrix
models of the two-dimensional gravity give us another important example [22].
The partition function for the one-matrix model in the double scale limit
coincides with the $\tau$-function, corresponding to a special KdV solution,
determined by the constraint ([22]):
$$
\frac {\partial u}{\partial \beta_{-2}} = 0.
\eqno (0.13)
$$
\par
In the conformal field theory the algebra of the vector fields in the circle
and its central extension - the Virasoro algebra play the key role.
A subalgebra of the algebra of nonisospectral symmetries corresponds to them
in the KP theory (the central extension does emerge if we calculate the
action of these symmetries at the $\tau$-function) (see [34], [20]) .
\par
Similar results for the two-dimensional topological quantum theory were
obtained by I.M.Krichever [23]. He proved that the averaged KP equations
are connected with the topological models and calculated the action of
the Virasoro symmetries on these equations in terms of the averaged
$\tau$-function. Applications of the Whitham equations to the topological
models were also considered by B.A.Dubrovin [6].
\par
In our paper we use the results of [23], [20], [21] to calculate the action
of nonisospectral symmetries on the averaged KdV in terms of the Riemann
invariants. We will assume the averaged KdV to be written in the Flaschka -
Forest - McLaughlin form. The plan of our parer in the following. In the
first section we recall the necessary definitions and results from the
Riemann surfaces theory, including the deformations of the complex structures
via the vector fields action. In the section 2 we recall the necessary results
from the periodic KdV and KP theory including the action of the nonisospectral
symmetries on the finite-gap solutions. The section 3 is dedicated to the
averaged KdV equations. We recall the construction of the Abel Whitham
hierarchy and present a new noncommutative set of symmetries.
\vspace{24.0pt}
\par
\noindent
{\bf {1. THE RIEMANN SURFACES THEORY. SOME DEFINITIONS AND RESULTS.}}
\vspace{12.0pt}
\par
In this section we recall the  constructions from the Riemann surfaces
theory we will use later ([24] - [26], [6], [20], [33]).
\par
In our paper the words "Riemann surface" will always denote a compact
nondegenerate Riemann surface. Such surfaces may be characterized as
close Riemann surfaces of finite genus or as algebraic Riemann surfaces.
We will consider two main classes - the general finite-gap surfaces
(they correspond to the KP quasiperiodic solutions) and the hyperelliptic
Riemann surfaces with a branch point in the infinity (they correspond
to the quasiperiodic KdV solutions).
\par
A Riemann surface $\Gamma$ is called hyperelliptic if there exists a
meromorphic function $E$ on $\Gamma$ with two simple poles or with one
second-order pole. This function maps $\Gamma$ to the complex plane,
the covering $E: \Gamma \rightarrow \bar {\bf C}$ is two-sheeted so
$\Gamma$ is isomorphic to
$$
\hbox {a)} \ \ Y^2=(E-E_1)\ldots (E-E_{2g+2}) \ \hbox {or}
\eqno (1.1.a)
$$
$$
\hbox {b)} \ \ Y^2=(E-E_1)\ldots (E-E_{2g+1})
\eqno (1.1.b)
$$
respectively (here $g$ is the genus of $\Gamma$). The branch points of $\Gamma$
over $E$ are $E_1$, \ldots, $E_{2g+2}$ in the case a or $E_1$, \ldots,
$E_{2g+1}$, $\infty$ in the case b. In our paper only the hyperelliptic
surfaces of the type b will be considered because only they are related to
the KdV theory.
\vspace {12pt}
\par
\noindent
{\bf 1.1. Cycles On The Riemann Surfaces.}
\vspace {12pt}
\par
Any compact nondegenerate Riemann surface is topologically equivalent to the
sphere with a finite number of handles. This number is called genus, we
will denote it $g$.
\par
The canonical basis of 1-cycles in $\Gamma$ consists of $2g$ elements $a_1$,
\ldots, $a_g$, $b_1$, \ldots, $b_g$ such, that
$$
a_i \circ a_j = b_i \circ b_j = 0, \ \ a_i \circ b_j = \delta _{ij}.
\eqno (1.2)
$$
where $\circ$ denotes the intersection number. (Of course there are infinitely
many nonhomotopical choices of canonical bases.) The surface $\Gamma$ can be
described as the result of gluing the sides of a $4g$-sided polygon (the
picture corresponds to $g=2$)
\par
\noindent
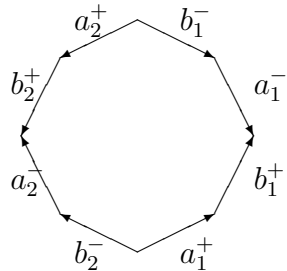
\setlength{\unitlength}{0.8pt}
\begin{picture}(400,150)
\put(200,20){\vector(2,1){36.6666}}
\put(200,20){\vector(-2,1){36.6666}}
\put(200,130){\vector(2,-1){36.6666}}
\put(200,130){\vector(-2,-1){36.6666}}
\put(236.6666,38.3333){\vector(1,2){18.3333}}
\put(163.3334,38.3333){\vector(-1,2){18.3333}}
\put(236.6666,111.6664){\vector(1,-2){18.3333}}
\put(163.3334,111.6664){\vector(-1,-2){18.3333}}
\put(220,15){$a^+_1$}
\put(255,50){$b^+_1$}
\put(255,95){$a^-_1$}
\put(220,125){$b^-_1$}
\put(170,15){$b^-_2$}
\put(140,50){$a^-_2$}
\put(140,95){$b^+_2$}
\put(170,125){$a^+_2$}
\end{picture}
\par
\noindent
in the following way: $a^+_i \leftrightarrow a^-_i$,
$b^+_i \leftrightarrow b^-_i$. (All the vertices are glued together). The
sides $a^{\pm}_i$, $b^{\pm}_i$ correspond to the cycles $a_i$, $b_i$
respectively.
\par
Let $\Gamma$ be a hyperelliptic surface - a two-sheeted covering over the
$E$-plane with the branch points $\infty$, $E_1$, \ldots, $E_{2g+1}$ and
the cuts $(-\infty ,E_1)$, $(E_2,E_3)$, \ldots, $(E_{2g},E_{2g+1})$. Then
we can choose the following canonical basis
\par
\noindent
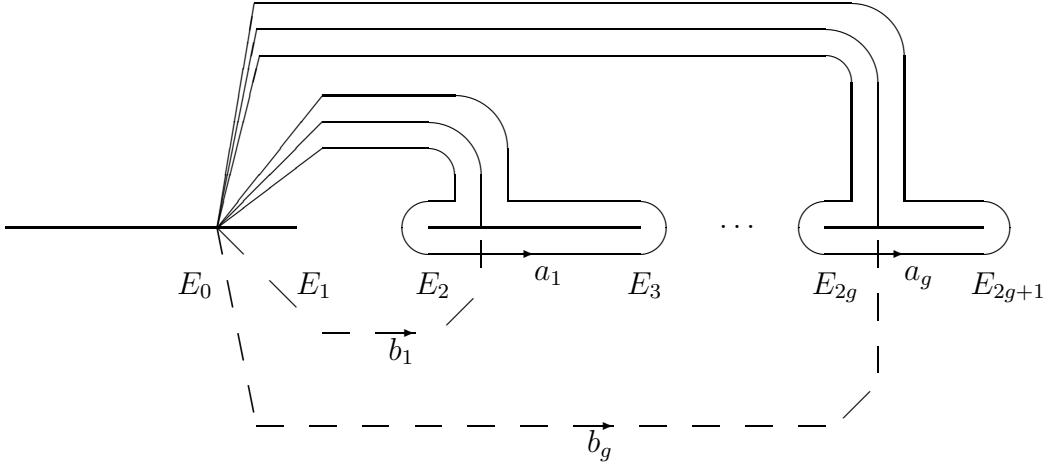
\setlength{\unitlength}{1pt}
\begin{picture}(410,200)
\put(0,100){\line(1,0){110}}
\put(80,100){\line(1,6){14.16666}}
\put(80,100){\line(1,5){15}}
\put(80,100){\line(1,4){16.25}}
\put(94.16666,185){\line(1,0){215.83334}}
\put(95,175){\line(1,0){215}}
\put(96.25,165){\line(1,0){213.57}}
\put(310,155){\oval(20,20)[tr]}
\put(310,155){\oval(40,40)[tr]}
\put(310,155){\oval(60,60)[tr]}
\put(320,110){\line(0,1){45}}
\put(330,100){\line(0,1){55}}
\put(340,110){\line(0,1){45}}
\put(80,100){\line(4,5){40}}
\put(80,100){\line(1,1){40}}
\put(80,100){\line(4,3){40}}
\put(120,150){\line(1,0){40}}
\put(120,140){\line(1,0){40}}
\put(120,130){\line(1,0){40}}
\put(160,120){\oval(20,20)[tr]}
\put(160,120){\oval(40,40)[tr]}
\put(160,120){\oval(60,60)[tr]}
\put(170,110){\line(0,1){10}}
\put(180,100){\line(0,1){20}}
\put(190,110){\line(0,1){10}}
\multiput(80,100)(4,-20){4}{\line(1,-5){2}}
\multiput(95,25)(20,0){6}{\line(1,0){10}}
\multiput(240,25)(20,0){4}{\line(1,0){10}}
\put(215,25){\vector(1,0){15}}
\multiput(330,95)(0,-20){3}{\line(0,-1){10}}
\put(317,28){\line(1,1){10}}
\multiput(80,100)(20,-20){2}{\line(1,-1){10}}
\put(120,60){\line(1,0){10}}
\put(140,60){\vector(1,0){15}}
\put(180,95){\line(0,-1){10}}
\put(167,63){\line(1,1){10}}
\put(160,100){\line(1,0){80}}
\put(160,100){\oval(20,20)[l]}
\put(240,100){\oval(20,20)[r]}
\put(160,110){\line(1,0){10}}
\put(190,110){\line(1,0){50}}
\put(160,90){\vector(1,0){40}}
\put(200,90){\line(1,0){40}}
\put(270,100){\ldots}
\put(310,100){\line(1,0){60}}
\put(310,100){\oval(20,20)[l]}
\put(370,100){\oval(20,20)[r]}
\put(310,110){\line(1,0){10}}
\put(340,110){\line(1,0){30}}
\put(310,90){\vector(1,0){30}}
\put(340,90){\line(1,0){30}}
\put(65,75){$E_0$}
\put(110,75){$E_1$}
\put(155,75){$E_2$}
\put(235,75){$E_3$}
\put(305,75){$E_{2g}$}
\put(365,75){$E_{2g+1}$}
\put(200,80){$a_1$}
\put(340,80){$a_g$}
\put(145,50){$b_1$}
\put(220,15){$b_g$}
\end{picture}
\par
\noindent
where the point $E_0$ correspond to the vertices of the polygon.
\vspace*{12pt}
\par
\noindent
{\bf 1.2. Differentials On The Riemann Surfaces.}
\vspace {12pt}
\par
Let us have a marked point $\infty$ in our Riemann surface $\Gamma$ with a
local parameter $z$. We use the notation $\infty$ because this point
corresponds to the infinite value of energy in the spectral theory. The
local parameter $z$ is a function defined in the neighbourhood of $\infty$
such, that $z(\infty)=0$, $dz(\gamma)\mid_{\gamma = \infty} \neq 0$. For
convenience we introduce an additional function $\lambda (\gamma)=1/z(\gamma)$.
For the hyperelliptic surfaces we will always assume $E(\infty ) = \infty $ and
$E=-\lambda ^2$. All the $\lambda$-expansions will be defined in the
neighbourhood of the point $\infty$.
\par
For us the following objects in $\Gamma$ will be necessary.
\par
1) Canonical basis of holomorphic differentials $\omega_1$, \ldots, $\omega_g$.
Differentials $\omega_j$ are determined by the normalization conditions
$$
\oint_{a_j} \omega_i = \delta_{ij}.
\eqno (1.3)
$$
The matrix $(B_{ij})$
$$
B_{ij} = \oint_{b_j} \omega_i.
\eqno (1.4)
$$
is called Riemann matrix. It is symmetrical $(B_{ij}=B_{ji})$, the imaginary
part of the Riemann matrix $\hbox {Im} \ B_{ij}$ is positive definite. The
coefficients of the Taylor series for $\omega_j$ in $\infty$ we will denote
$q^{H}_{jk}$.
$$
\omega_j=\sum_{m \geq 1} q^H_{jm} \lambda^{-m-1} d \lambda
\eqno (1.5)
$$
\par
2) Meromorphic differentials $\Omega_j$ with the only pole in the point
$\infty$ such, that
$$
\Omega_k = d (\lambda ^ k) + O(1), \ \ \oint_{a_i} \Omega _k = 0.
\eqno (1.6)
$$
\par
We denote
$$
\oint_{b_j} \Omega _k = (U_k)_j, \ {\vec U}_k=((U_k)_1, \ldots , (U_k)_g)
$$
$$
\Omega_k = d (\lambda ^ k) + \sum _{m \geq 1} Q_{km} \lambda^ {-m-1} d \lambda.
\eqno (1.7)
$$
\par
The Riemann bilinear relations (see (1.16) below) result in
$$
(U_j)_k = - 2 \pi i q^H_{kj} ,\ \ Q_{ij}=Q_{ji}.
\eqno (1.8)
$$
If $\Gamma$ is a hyperelliptic curve then $\Omega_{2k} = d ((-E)^k)$,
$(U_{2k})_l=0$, $Q_{kl}=0$ if at least one of the indexes $k$, $l$ is even.
\par
3) Multivalued holomorphic differentials $\omega^i_k$, $\sigma ^i_k$ (they are
defined {\em {only}} in the {\em {hyperelliptic}} Riemann
surfaces). These differentials are determined by the following properties
$$
\Delta_{a_j} \omega ^i_k = \delta_{ij} d(E^k), \
\Delta_{b_j} \omega ^i_k = 0,
\eqno (1.9)
$$
$$
\Delta_{b_j} \sigma ^i_k = - \delta_{ij} d(E^k), \
\Delta_{a_j} \sigma ^i_k = 0,
\eqno (1.10)
$$
$$
\oint_{a_j} \omega^i_k = \oint_{a_j} \sigma^i_k =0.
\eqno (1.11)
$$
Here $\Delta_{a_i}$ and $\Delta_{b_i}$ mean the increment of the differentials
when going along the cycles $a_i$ and $b_i$ respectively. We will denote
$$
\oint_{b_j} \omega^i_k = (U^i_k)_j, \
\oint_{b_j} \sigma^i_k = (V^i_k)_j
\eqno (1.12)
$$
$$
\omega^i_k = \sum_{m \geq 1} Q^i_{km} \lambda^{-m-1} d \lambda, \
\sigma^i_k = \sum_{m \geq 1} R^i_{km} \lambda^{-m-1} d \lambda,
\eqno (1.13)
$$
\par
4) Quasimomentum 1-differential $dp$, meromorphic in $\Gamma$ with the only
pole in the point $\infty$ such, that
$$
dp=-i d(\lambda)+O(1), \
\hbox{Im}\ \oint_{a_j} dp = \hbox{Im}\ \oint_{b_j} dp =0 .
$$
\par
5) Algebra of the holomorphic vector fields in the punctured neighbourhood
of $\infty$ with the basis
$$
l_i = \lambda^{i+1} \partial _\lambda,  -\infty < i < \infty.
\eqno (1.14)
$$
These fields have the following commutators
$$
[l_i,l_j]=(j-i) l_{i+j}.
\eqno (1.15)
$$
\par
In the hyperelliptic case we will consider a subalgebra, generated by
the elements with even indexes. All elements of this subalgebra are
single-valued in the $E$-plane (in the neighbourhood of infinity).
$$
L_i=l_{2i}=2 (-1)^i E^{i+1} \partial _E.
$$
\par
The deformations of $\Gamma$, generated by these elements preserve the
hyperelliptic structure. (The action of the vector fields on the Riemann
surfaces will be described below in the section 1.3).
\par
6) Holomorphic 2-differentials in $\Gamma$. These differentials can be
written in the local coordinates as $\omega^{(2)} = \omega^{(2)}(z) (dz)^2$,
the space of such differentials is $3g-3$ - dimensional as $g>1$,
1 - dimensional as $g=1$ and empty as $g=0$. We have a standard overdetermined
full set in this space $\omega^{(2)}_{ij}=\omega_i \omega_j$ where $\omega_i$
are holomorphic 1-differentials.
\par
7) Riemann bilinear relations.
\par
Let $\Omega_1$, $\Omega_2$ be multivalued meromorphic differentials in $\Gamma$
such, that
\par
a) $\Omega_1$, $\Omega_2$ have no branch points and are locally
single-valued.
\par
b) All the residues of $\Omega_1$, $\Omega_2$ are equal to 0.
\par
c) $d^{-1} \Delta_{a_i} \Omega_j$, $d^{-1} \Delta_{b_i} \Omega_j$ are
single-valued meromorphic functions in $\Gamma$ without singularities on
the cycles. We assume all these functions to be equal to 0 in the point
of intersection of all cycles (this point correspond to the vertices of
the polygon).
\par
Then we have the following relation:
$$
2 \pi i \sum \hbox{res} (d^{-1} \Omega _1) \Omega _2 = \sum _i
\left \{ \oint_{a_i} \Omega_1 \oint_{b_i} \Omega_2-
\oint_{a_i} \Omega_2 \oint_{b_i} \Omega_1 +
\oint_{a_i} (d^{-1} \Delta_{b_i} \Omega_2) \Omega_1 \right \} +
$$
$$
+ \sum _i \left \{ - \oint_{a_i} (d^{-1} \Delta_{b_i} \Omega_1) \Omega_2  +
\oint_{b_i} (d^{-1} \Delta_{a_i} \Omega_1) \Omega_2 -
\oint_{b_i} (d^{-1} \Delta_{a_i} \Omega_2) \Omega_1 \right \} +
\eqno (1.16)
$$
$$
+ \sum _i \left \{ - \oint_{a_i} (d^{-1} \Delta_{b_i} \Omega_1)
(\Delta_{b_i} \Omega_2) -
\oint_{b_i} (d^{-1} \Delta_{a_i} \Omega_2) (\Delta_{a_i} \Omega_1) \right \} .
$$
\par
All the terms in the right-hand side of (1.16) are correctly defined because
of c). The property b) guaranties the correctness of the left-hand side.
\par
{\bf Remark.} All the residues of $\Omega_1$, $\Omega_2$ are equal to 0 so
we have
$$
\sum \hbox{res} (d^{-1} \Omega _1) \Omega _2 =
- \sum \hbox{res} (d^{-1} \Omega _2) \Omega _1
\eqno (1.17)
$$
\par
8) We will use the following scalar product, introduced in [27] and
generalized in [23], [6]. Let the 1 - forms $\Omega_1$, $\Omega_2$ be
linear combinations of differentials, defined in the points 1-3. Then
$$
V_{\Omega_1 \Omega_2} = \hbox{res}\ (d^{-1} \Omega_1)_{+} \Omega_2 -
\frac{1}{2 \pi i} \sum_k \oint_{a_k} \Omega_1 \oint_{b_k} \Omega_2 +
$$
$$
+ \frac{1}{2 \pi i} \sum_k \oint_{a_k} (d^{-1} \Delta _{b_k} \Omega_1)
\Omega_2 -
\frac{1}{2 \pi i} \sum_k \oint_{b_k} (d^{-1} \Delta _{a_k} \Omega_1) \Omega_2.
\eqno (1.18)$$
(We have only one singular point in (1.18) so we omit the $\sum$ sign
before the residue).
The brackets $(\ \ )_+$ in (1.18) denote the singular part of a function
$$
\left ( \sum_{m=-\infty}^{m=+\infty} a_m \lambda^m \right ) _+ =
\sum_{m=1}^{m=+\infty} a_m \lambda^m
\eqno (1.19)
$$
\par
Comparing (1.17) with the Riemann relations (1.16) we see that the product
(1.18) is symmetrical
$$
V_{\Omega_1 \Omega_2} = V_{\Omega_2 \Omega_1}.
\eqno (1.20)
$$
( The last two terms in (1.16) are equal to 0. )
\par
Simple direct calculations result in:
$$
V_{\Omega_k \Omega_l} = Q_{kl} = Q_{lk}
$$
$$
V_{\omega_k \Omega_l} = q^H_{kl}
$$
$$
V_{\omega^\alpha_k \Omega_l} = Q^{\alpha}_{kl} = - \frac {1}{2 \pi i}
\oint_{b_\alpha} (E^k -E_0^k ) \Omega _l
$$
$$
V_{\sigma^\alpha_k \Omega_l} = R^{\alpha}_{kl} = - \frac {1}{2 \pi i}
\oint_{a_\alpha} (E^k -E_0^k ) \Omega _l
$$
$$
V_{\omega_k \omega_l} = - \frac {1}{2 \pi i}  B_{ik}
$$
$$
V_{\omega_k \omega^\alpha_l} = - \frac{1}{2 \pi i} (U^\alpha_l)_k=
- \frac {1}{2 \pi i} \oint_{b_\alpha} (E^l -E_0^l ) \omega _k
\eqno (1.21)
$$
$$
V_{\omega_k \sigma^\alpha_l} = - \frac{1}{2 \pi i} (V^\alpha_l)_k=
- \frac {1}{2 \pi i} \oint_{a_\alpha} (E^l -E_0^l ) \omega _k
$$
$$
V_{\omega^\alpha_k \omega^\beta_l} =
- \frac {1}{2 \pi i} \oint_{b_\alpha} (E^k -E_0^k ) \omega^\beta_l
$$
$$
V_{\omega^\alpha_k \sigma^\beta_l} =
- \frac {1}{2 \pi i} \oint_{b_\alpha} (E^k -E_0^k ) \sigma^\beta_l
$$
$$
V_{\sigma^\alpha_k \sigma^\beta_l} =
- \frac {1}{2 \pi i} \oint_{a_\alpha} (E^k -E_0^k ) \sigma^\beta_l
$$
\par
The following property of the scalar product $V_{\Omega_1 \Omega_2}$ is
important for the averaged equations theory.
\par
{\bf Lemma 1.1} Let $\Gamma$ be a hyperelliptic Riemann surface with the
branch points $\infty$, $E_1$, \ldots, $E_{2g+1}$. Then the scalar products
$V_{\Omega_k \Omega_l}$, $V_{\omega_k \Omega_l}$,
$V_{\omega^\alpha_k \Omega_l}$, $V_{\sigma^\alpha_k \Omega_l}$,
$V_{\omega_k \omega_l}$, $V_{\omega_k \omega^\alpha_l}$,
$V_{\omega_k \sigma^\alpha_l}$ can be expressed via branch points
$E_1$, \ldots, $E_{2g+1}$ and the normalization point $E_0$ in terms of the
hyperelliptic integrals (elliptic integrals as $g=1$).
\par
Proof of the Lemma. The forms $\omega_k$, $\Omega_k$ in the hyperelliptic case
have the following representations
$$
\omega_k = \frac {\left ( \sum_{l=1}^g A_{kl}
E^{l-1} \right ) dE}{\sqrt {-4(E-E_1) \ldots (E-E_{2g+1})}}
$$
$$
\Omega_k = \frac {\left ( -k (-E)^{g+\frac{k-1}{2}} + \sum_{l=1}^g C_{kl}
E^{l-1} \right ) dE}{\sqrt {-4(E-E_1) \ldots (E-E_{2g+1})}},
\eqno (1.22)
$$
$$ k=2n+1, n \ge 0. $$
\par
The coefficients $A_{kl}$, $C_{kl}$ are uniquely determined by the
normalization conditions
$$
\oint_{a_j} \omega_i = \delta_{ij},\ \ \oint_{a_j} \Omega_i = 0.
\eqno (1.23)
$$
\par
Equations (1.23) are equivalent to a linear system on $A_{kl}$, $C_{kl}$.
All the coefficients of this system are hyperelliptic integrals, the system
is non-degenerate (see [25]) and we can solve it. The coefficients of the
Taylor series in $\infty$ for $\omega_k$, $\Omega_k$ can be algebraically
expressed via $A_{kl}$, $C_{kl}$, $E_j$. The remark that all the integrals
$$
\oint (E^k -E_0^k ) \omega _l,\ \ \oint (E^k -E_0^k ) \Omega _l
\eqno (1.24)
$$
are hyperelliptic completes the proof.
\vspace {12pt}
\par
\noindent
{\bf 1.3. Deformations Of The Riemann Surfaces And The Riemann Problem.
The Moduli Space.}
\vspace {12pt}
\par
All the surfaces of genus $g$ are topologically equivalent. But if $g \geq 1$
then they may be different as complex manifolds. Consider the simplest case
$g=1$ of complex tori. There exists an unique holomorphic differential
$\omega_1$ such that
$$
\oint_a \omega_1 = 1.
\eqno (1.25)
$$
\par
The parameter
$$
\tau = \oint_b \omega_1 , \ \ \hbox{Im} \ \tau > 0
\eqno (1.26)
$$
is correctly defined if we have a fixed basis of 1 - cycles $a$, $b$ in
$\Gamma$. The surface $\Gamma$ can be described as a factor of the complex
plane by the group of shifts, generated by:
$$
z \rightarrow z+1, z \rightarrow z+\tau.
\eqno (1.27)
$$
\par
Let $\Gamma$, $\Gamma '$ be complex tori (g=1) with the bases of
1 - cycles $a$, $b$ and $a'$, $b'$ respectively. Then a complex map
$f:\Gamma \rightarrow \Gamma'$ such that $f(a)=a'$, $f(b)=b'$
exists if and only if $\tau =\tau'$.
\par
Parameter $\tau$ depends on the choice of basic cycles. Let us have two
bases $a$, $b$ and $\tilde{a}$, $\tilde{b}$ respectively. Then there exists
a matrix $\left ( \begin{array}{cc} \alpha & \beta \\ \gamma & \delta
\end{array} \right )
\in sl(2,{\bf Z})$ such, that
$$
\left (\begin{array}{c} \tilde{a} \\ \tilde{b} \end{array} \right ) =
\left ( \begin{array}{cc} \alpha & \beta \\ \gamma & \delta \end{array}
\right )
\left ( \begin{array}{c} a \\ b \end{array} \right ),\ \
\tilde{\tau} = \frac {\gamma + \delta \tau} {\alpha + \beta \tau}.
\eqno (1.28)
$$
\par
The group $sl(2,{\bf Z})$ consists of all $2 \times 2$ integer matrices such
that $\alpha \delta - \beta \gamma = 1$.
\par
Summing all of this we have:
\par
Two complex (one-dimensional in the complex sense) tori with the parameters
$\tau$ and $\tau'$ are isomorphic if and only if there exists a matrix
$$
\left ( \begin{array}{cc} \alpha & \beta \\ \gamma & \delta \end{array}
\right )
\in sl(2,{\bf Z}) \ \ \hbox{such, \ that} \ \
\tau' = \frac {\gamma + \delta \tau} {\alpha + \beta \tau}.
\eqno (1.29)
$$
The space of all 1-dimensional complex tori is called the moduli space for
$g=1$ or the space of all complex structures on the surfaces of genus $g=1$.
We have proved that the moduli space for $g=1$ is the factor of the complex
upper-plane $\hbox{Im}\ \tau > 0$ by the group $sl(2,{\bf Z})$.
\par
It is possible to consider the moduli space of all Riemann surfaces with a set
of marked points and the moduli space of the Riemann surfaces with a set of
marked points and a local parameter in one of them. The last space is
infinite-dimensional.
\par
The moduli space of the Riemann surfaces with $g>1$ is a $3g-3$-dimensional
complex manifold. We do not want to discuss its properties in details. But
the following construction is very important for us.
\par
Let $\Gamma$ be a Riemann surface with a marked point $\infty$, a local
parameter $z=1/\lambda$ in the neighbourhood of $\infty$, $S$ be a small
contour surrounding $\infty$, $v=v(\lambda) d \lambda$ be a holomorphic
vector field in the vicinity of the contour $U(S)$. Then we can construct
a family of new Riemann surfaces $\Gamma_\alpha$ depending on a parameter
$\alpha$ (see [25]). We will assume $\alpha$ to be sufficiently small.
\par
The contour $S$ splits $\Gamma$ to a small disk $D$, containing the point
$\infty$ and an open Riemann surface $\Gamma \backslash D$ (
$\Gamma \backslash D$ denotes the set of point $\gamma$ such, that
$\gamma \in \Gamma$, $\gamma \not \in D$). We can cover $\Gamma$ by two
regions $\Gamma_+$, $\Gamma_-$ such, that:
\par
1) $D \subset \Gamma_-$.
\par
2) $\Gamma \backslash D \subset \Gamma_+$.
\par
3) $\Gamma_+ \cap \Gamma_- = U(S)$.
\par
If $\gamma$ is a point of $U(S)$ then we will denote the correspondent points
in $\Gamma_+$ and $\Gamma_-$ $\gamma_+$ and $\gamma_-$ respectively.
$\Gamma$ may be treated as the result of gluing $\Gamma_+$ to $\Gamma_-$
$f: \Gamma_+ \rightarrow \Gamma_-$ \ $f(\gamma_+) = \gamma_-$. Let us
introduce a new gluing function $f_\alpha$ by formula
$$
f_\alpha (\gamma_+) = \exp (\alpha v) \gamma_-,
\eqno (1.30)
$$
where $\exp (\alpha v) \gamma_-$ denotes the shift of the point $\gamma_-$ via
the vector field $v$ after the lapse of the time $\alpha$. Then we can define
a new Riemann surface $\Gamma_\alpha$ as the result of gluing
$\Gamma_+$ to $\Gamma_-$ via the function $f_\alpha$.
\par
The Riemann surfaces $\Gamma$ and $\Gamma_\alpha$ are constructed from the same
parts $\Gamma_+$ and $\Gamma_-$. Thus we have a map ${\bf E}:
\Gamma \rightarrow \Gamma_\alpha$, coinciding with identical maps
$\Gamma_+ \rightarrow \Gamma_+$ on $\Gamma \backslash D$ and
$\Gamma_- \rightarrow \Gamma_-$ on $D$. Map ${\bf E}$ has a jump on the contour
$S$. If we have some marked points in $\Gamma$ or local parameters in some of
them then we will map these objects by {\bf E}.
\par
Let us have a infinitesimal transformation ($\alpha$ is infinitely small) and
$\Delta$ be a holomorphic tensor field in $\Gamma_\alpha$. The map ${\bf E}$
carries $\Delta$ from $\Gamma_\alpha$ to $\Gamma$ so we can treat $\Delta$ as
a tensor field in $\Gamma$ with a jump on the contour $S$. The boundary values
of $\Delta$ on $S$ \ $\Delta_+$ and $\Delta_-$ satisfy the following relation
$$
\Delta_+ - \Delta_- = \alpha L_v \Delta,
\eqno (1.31)
$$
where $L_v$ denotes the Lie derivative
$$
L_{v(\lambda) \partial_\lambda} \ g(\lambda) (d \lambda)^\alpha =
(v(\lambda) g'(\lambda) + \alpha v'(\lambda) g(\lambda))\ (d \lambda)^\alpha
\eqno (1.32)
$$
($\alpha$ is infinitely small so we can write $\Delta_+$ in the right-hand
side of (1.31) as well as $\Delta_-$).
\par
Formula (1.31) gives us a very convenient method for calculation the variations
of tensor object via complex structure variations: we calculate the right-hand
side of (1.31) and solve the Riemann problem in the appropriate functional
class. Riemann problem is one of the basic objects in the soliton theory so
this approach allows us to connect these deformations of the Riemann surfaces
with the non-isospectral symmetries (see section 2.2).
\par
In the hyperelliptic case our construction has the following interpretation.
Let $\Gamma$ be a hyperelliptic surface over the $E$-plane with the branch
points $\infty$, $E_1$, \ldots, $E_{2g+1}$, $v=v(E)\partial _E$ be a vector
field in the $E$-Plane, nonsingular for all $E \neq \infty$, the local
parameter in the point $\infty$ be $z=1/\lambda$, $\lambda=\sqrt{-E}$.
\par
Then $\Gamma_\alpha$ is a hyperelliptic Riemann surface with the branch
points $\infty$, $\exp (\alpha v) E_1$, \ldots, $\exp (\alpha v) E_{2g+1}$
and the local parameter $z=1/\lambda$, $\lambda=\sqrt{-E}$. The map ${\bf E}$
carries the point $E_{\pm}$ in $\Gamma$ to the point $\exp (\alpha v) E_{\pm}$
in $\Gamma_\alpha$ if the point $E_{\pm}$ is located outside the neighbourhood
of $\infty$ surrounded by a small contour $S$ and carries the point $E_{\pm}$
in $\Gamma$ to the point $E_{\pm}$ in $\Gamma_\alpha$ if the point $E_{\pm}$
is located inside the neighbourhood of $\infty$ (the sign $\pm$ means the
upper or lower sheet).
\par
If we have a family of deformed hyperelliptic Riemann surfaces $\Gamma_\alpha$
and a function on this family $f(E_\pm,\alpha)$ then we have two differential
operators
$$
\partial_\alpha f = \frac{\partial f}{\partial \alpha}
\eqno (1.33)
$$
and
$$
D_\alpha f = \frac{\partial f}{\partial \alpha} + L_v f,
\eqno (1.34)
$$
where $L_v$ is the Lie derivative (1.32) in the first argument. In contrast
with $\partial_\alpha$ the operator $D_\alpha$ can be generalized to arbitrary
Riemann surfaces and can be treated as a connection generated by the map $E$.
\par
The following property of the connection $D_\alpha$ is very important for us:
\par
{\bf Lemma 1.2.} Let $\Delta(E_\pm,\alpha)$ be a holomorphic tensor field
in $\Gamma_\alpha \backslash \infty$ for all $\alpha$. Then
$D_\alpha \Delta(E_\pm,\alpha)$ is nonsingular outside the neighbourhood of
$\infty$ (for $\partial_\alpha$ in is not valid in the branch points).
\par
While calculating the nonisospectral symmetries we will need the derivatives
of the holomorphic differentials by the complex structures. In the
hyperelliptic case we have:
\par
{\bf Lemma 1.3.} Let $\Gamma$ be a hyperelliptic Riemann surface,
$L_n=l_{2n}=2 (-1)^n E^{n+1} \partial _E$ be a holomorphic vector field in
the $E$-plane, $i \geq -1$, $\Gamma_\alpha$ be the deformation of $\Gamma$ via
the vector field $v$. Then
$$
D_\alpha \Omega_k = k \Omega_{2n+k} + \sum_{m=1}^{2n-1} Q_{km} \Omega_{2n-m},
\eqno(1.35)
$$
$$
D_\alpha \omega_k = \sum_{m=1}^{2n-1} q^H_{km} \Omega_{2n-m},
\eqno(1.36)
$$
$$
D_\alpha \omega^\alpha_k = 2 (-1)^n k \omega^\alpha_{n+k} +
\sum_{m=1}^{2n-1} Q^\alpha_{km} \Omega_{2n-m},
\eqno(1.37)
$$
$$
D_\alpha \sigma^\alpha_k = 2 (-1)^n k \sigma^\alpha_{n+k} +
\sum_{m=1}^{2n-1} R^\alpha_{km} \Omega_{2n-m},
\eqno(1.38)
$$
\par
Proof of the Lemma 1.3. In accord with the rule described above we have
to solve the following Riemann problem:
\par
Let $\Delta$ be one of the differentials $\Omega_k$, $\omega_k$,
$\omega^\alpha_k$, $\sigma^\alpha_k$. Then we have to construct a pair
of differentials $\Delta_+$, $\Delta_-$ such that
\par
1) $\Delta_+ - \Delta_- = L_v \Delta$.
\par
2) $\Delta_-$ is defined and nonsingular in the neighbourhood of $\infty$.
\par
3) $\Delta_+$ is nonsingular for all $E_{\pm} \neq \infty$.
\par
4) $\oint_{a_i} \Delta_+ = 0$.
\par
5) $\Delta_+$ is single-valued as $\Delta = \Omega_k$ or $\Delta = \omega_k$,
$$
\Delta_{a_i} \Delta_+ = 2 (-1)^n k \Delta_{a_i} \omega^\alpha_{n+k}, \
\Delta_{b_i} \Delta_+ = 0 \ \hbox{as}\ \Delta_+ = \omega^\alpha_k
$$
$$
\Delta_{b_i} \Delta_+ = 2 (-1)^n k \Delta_{a_i} \sigma^\alpha_{n+k}, \
\Delta_{a_i} \Delta_+ = 0 \ \hbox{as}\ \Delta_+ = \sigma^\alpha_k
$$
\par
Then $D_\alpha \Delta = \Delta_+$.
\par
In fact we are looking for a differential with the properties 3-5 in the
finite part of $\Gamma$ and the prescribed singularity $L_v \Delta$ in
$\infty$. But such differential can be easily constructed as a linear
combination of $\Omega_k$, $\omega_k$, $\omega^\alpha_k$, $\sigma^\alpha_k$.
This completes the proof.
\par
{\bf Corollary 1.} ([25]).
$$
\frac {\partial B_{ij}}{\partial \alpha} = \oint_{S} v \omega_i \omega_j.
\eqno (1.39)
$$
where the product $v(\lambda) \partial_\lambda\ \omega_i(\lambda) d \lambda
\ \omega_j(\lambda) d \lambda$ is the 1-form $v(\lambda) \omega_i(\lambda)
\omega_j(\lambda) d \lambda$. This formula is valid for general Riemann
surfaces as well as for hyperelliptic ones. We assume that the contour $S$
goes around the point $\infty$ counterclockwise.
\par
{\bf Corollary 2.} The vector field $v$ do not vary the complex structure of
$\Gamma$ if and only if all the integrals
$$
\oint_{S} v \omega^{(2)}_i
$$
is are equal to 0. Here $\omega^{(2)}_i$ is the basis of holomorphic 2-forms.
It proves that the moduli space is $3g-3$ dimensional as $g>1$ and
1 - dimensional as $g=1$.
\vspace {12pt}
\par
\noindent
{\bf 1.4. Riemann Theta-Functions.}
\vspace {12pt}
\par
Let $b_{ij}$ be a complex symmetrical $g \times g$ matrix such, that
$\hbox{Re}\ b_{ij}$ is negative definite, $\vec z$ be a complex $g$-component
vector. Then the Riemann theta-function can be defined as an infinite sum
([26])
$$
\theta(\vec z|b_{ij}) = \sum_{m_1, \ldots, m_g} \exp \left \{ \frac{1}{2}
\sum_{kj} b_{kj}m_k m_j + \sum_k z_k m_k \right \} ,
\eqno (1.40)
$$
where $m_k$, $k=1, \ldots,g$ are arbitrary integers.  This sum converges
for all $\vec z$. The theta function has the following periodicity properties
$$
\theta(z_1,z_2, \ldots, z_k+2 \pi i ,\ldots,z_g |b_{ij}) =
\theta(z_1,z_2, \ldots, z_k,\ldots,z_g |b_{ij}).
\eqno (1.41)
$$
$$
\theta(z_1+b_{1k},z_2+b_{2k}, \ldots, z_g+b_{gk}|b_{ij}) =
\theta(z_1,z_2, \ldots, z_g |b_{ij}) \exp \{-b_{kk}/2 -z_k\}.
\eqno (1.42)
$$
\par
The zeros ot the theta function are described by the following Lemma (see [26])
\par
{\bf Lemma 1.4.} Let $\Gamma$ be a Riemann surface with a marked point
$\gamma_0$, $B_{ij}$ be the matrix of periods (1.4),
$b_{ji}=2 \pi i B_{ji}$, $\vec {A}(\gamma)$ be the Abel transform, i.e.
$\vec {A}(\gamma)$ is a multivalued map $\Gamma \rightarrow {\bf C}^g$
determined by the formula
$$
({\vec A}(\gamma))_k = 2 \pi i \int_{\gamma_0}^\gamma \omega_k, \ \
k=1, \ldots , g.
\eqno (1.43)
$$
\par
Then there exists a $g$-dimensional complex vector
$\vec K = \vec K (\Gamma,\gamma_0)$ such that the function
$$
\varphi (\gamma) = \theta ({\vec A}(\gamma) - {\vec A}(\gamma_1) - \ldots -
{\vec A}(\gamma_g) + {\vec K}|b_{ij}), \ \gamma \in \Gamma.
\eqno (1.44)
$$
has one of the following properties:
\par
1) $\varphi (\gamma) \equiv 0$ \ or
\par
2) $\varphi (\gamma)$ has exactly $g$ zeros in the points $\gamma_1$, \ldots,
$\gamma_g$.
\par
\noindent
$\vec K$ is called Riemann constants vector.
\vspace{24.0pt}
\par
\noindent
{\bf {2. PERIODIC THEORY OF THE KORTEVEG - DE - VRIES AND KADOMTSEV -
PETVIASHVILI EQUATIONS.}}
\vspace{12.0pt}
\par
In this section we will recall some facts from the KdV and KP theory. Two
topics are the most interesting for us: periodic (quasiperiodic) finite-gap
theory and the action of the nonisospectral symmetries on the finite-gap
solutions.
\par
Finite-gap KdV solutions can be treated as nonlinear superpositions of the
moving waves. Such solutions are the basic objects for the averaging
procedure. The have been constructed in the papers of S.P.Novikov,
B.A.Dubrovin, V.B.Matveev, A.R.Its, P.Lax, H.McKean and P. van Moerbeke
(see book [28] for more detailed description and references). Finite-gap
KP solutions were constructed by I.M.Krichever [29]. The direct periodic
problem for KP is more complicated (the results, obtained by I.M.Krichever
can be found in [29]).
\par
The first step in constructing averaged nonisospectral symmetries is the
following: the action of these symmetries on the finite-gap solution is
calculated. This problem was solved by A.Yu.Orlov and the author in [20],
[21]. We recall some results of these parers. We use the representation
for nonisospectral symmetries, suggested by A.Yu.Orlov and E.I.Schulman [17]
as the most convenient for us.
\vspace {12pt}
\par
\noindent
{\bf 2.1. KdV And KP Theory. Integration And Isospectral Symmetries. Periodic
Theory. Baker-Akhiezer Function And Cauchy Kernel.}
\vspace {12pt}
\par
The theory of the KdV equation
$$
u_t= \frac {1}{4} u_{xxx}-\frac{3}{2} uu_x.
\eqno (2.1)
$$
is based upon the existence of the following representation. Let
$$
L=-\partial _x^2+u(x,t),\
A=\partial _x^3-\frac {3}{4}(u\partial _x+\partial _x u)
\eqno(2.2)
$$
be ordinary differential operators depending on an extra parameter $t$. Then
the function $u(x,T)$ satisfy (2.1) if and only if the following relation takes
place
$$
\partial L/\partial t=[L,A].
\eqno(2.3)
$$
Representation (2.3) is called Lax pair or $L-A$ pair for KdV.
\par
One of the first results of the soliton theory was the existence of infinitely
many mutually commuting KdV symmetries. They can be written via the recursion
operator
$$
\frac {\partial u}{\partial t_{2n+1}}=\frac {d}{dx}
{\left(-\frac {\Lambda}{4}\right)}^n
u = -2\frac{d}{dx}{\left(-\frac {\Lambda}{4}\right)}^{n+1} \cdot 1,
\eqno (2.4)
$$
where
$$
\Lambda=-\partial _x^2+2\partial_x^{-1} u \partial_x +2u
\eqno (2.5)
$$
or in the Lax form
$$
\partial L/\partial_{t_{2n+1}}=[L,A_{2n+1}],
\eqno (2.6)
$$
where
$$
A_{2n+1}=\{ (-L)^{\frac{2n+1}{2}} \}_+,
\eqno (2.7)
$$
$(-L)^{\frac{2n+1}{2}}$ denotes a formal pseudodifferential operator, i.e a
series in $\partial _x$ with finite number of positive terms and infinite
number of negative, $\{\ \ \}_+$ denotes the differential part (see [31]).
We mark the times by odd indexes to have unified notations for KdV and KP.
Here $t_1=x$, $t_3=t$.
\par
Equations (2.3), (2.6) result in the following property: the spectrum of
L does not depend upon the times $t_3$, $t_5$, \ldots, $t_{2k+1}$, \ldots.
This is the reason why we call these symmetries isospectral.
\par
KdV equation is integrated by the inverse scattering transform, i.e. we
consider the "scattering data" for $L$ as a new variable instead of $u(x)$.
(We write "scattering data" in quotation marks to stress that these data
coincides with the physical scattering data only for some functional classes
of potentials). Because of the isospectral property the evolution law for the
"scattering data" is very simple (see [28]). The map from $u$ to the
scattering data for the vanishing in the infinity potentials in the small
amplitude limit coincides with the Fourier transform and can be treated as
its nonlinear analog.
\par
The realization of this scheme depends on the functional class of the
potential. Let us recall the scheme for the periodic case
$$
u(x+\Pi,t)=u(x,t),
\eqno (2.8)
$$
where $u(x,t)$ is a real nonsingular potential.
\par
We consider  the following spectral problems for $L$:
\par
a) Main problem $L\psi =E\psi$, $\psi$ is bounded in $x$.
\par
b) Auxiliary problem $L\psi _a=E\psi _a,\ \psi _a(0)=\psi _a(\Pi )=0$.
\par
The spectrum of the main problem consists of a set of intervals $[E_1,E_2]$,
\ $[E_3,E_4]$, \ldots ,\ $[E_{2n-1},E_{2n}]$, \ldots where
$E_1<E_2\le E_3<E_4\le E_5\ldots $,
$E_{2n+1}-E_{2n}\rightarrow 0$ as $n\rightarrow \infty$.
\par
The spectrum of the auxiliary problem consists of an infinite number of points
$d_1<d_2<d_3<\ldots $ located in the gaps $d_1\in [E_2,E_3],\ d_2\in [E_4,E_5],
\ d_3\in [E_6,E_7],\ldots$.
\par
The Bloch eigenfunction $\psi (x,E)$ normalized by the conditions $\psi(
x+\Pi ,E)=\exp (\Pi ip(E))$ $\psi (x,E)$ and $\psi (0,E)=1$ is meromorphic
on a two-sheeted Riemann surface $\Gamma $ over the E-plane with branch
points $E_1$, $E_2$,\ldots ,$\infty$, and has simple poles in the points
$\gamma_1$, $\gamma_2$, $\ldots $, such that the projection of $\gamma_n$
to the E-plane coincides with $d_n$.
\par
The function $p(E_{\pm})$ is defined in $\Gamma$ and is called quasimomentum.
\par
The spectrum corresponding to a general potential has infinite number of gaps.
But for us the so-called finite-gap case when $E_{2n}=E_{2n+1}$ for all $n>g$
is the most important. (The infinite-genus case can also be studied [32] but
the answers are much more complicated). Finite-gap potentials are dence in
the space of all periodic potentials.
\par
The inverse problem data in the finite-gap case is the following:
\par
1) $2g+1$ real numbers $E_1$, $E_2$, \ldots, $E_{2g+1}$,
$E_1<E_2<\ldots<E_{2g+1}$. The points $E_k$ are the boundary point of the
spectrum.
\par
2) $g$ points $\gamma_1$, \ldots, $\gamma_g$ in a hyperelliptic Riemann
surface $\Gamma$ with the branch points $\infty$, $E_1$, $E_2$, \ldots,
$E_{2g+1}$, such, that $E(\gamma_k) \in [E_{2k},E_{2k+1}]$ where $E$ is
the projection to the $E$-plane.
\par
This data uniquely determines the potential $u(x)$.
\par
The flows (2.3), (2.6) do not move the branch points $E_k$ but they shift
the divisor $\gamma_k$. The evolution of the points $\gamma_k$ can be
described in terms of ordinary differential equations, derived by
B.A.Dubrovin (see [28]). This system is nonlinear, but it has a very nice
explicit solution.
\par
{\bf Lemma 2.1.} (see [28]). Let ${\vec A}(\gamma)$ be the Abel transform
defined in the paragraph 1.4. Then
$$
\vec A(\gamma_1)+\vec A(\gamma_2)+\ldots+\vec A(\gamma_g)=\vec A_0 +
x \vec U_1 + t \vec U_3 + t_5 \vec U_5 + \ldots,
\eqno (2.9)
$$
where $\vec A_0$ is some constant vector, the vectors $U_k$ are defined by
(1.7).
\par
Inverse transform to (2.9) can be written in terms of the theta-functions.
The answer is given by the A.R.Its - V.B.Matveev formula (see [28], [33]).
$$
u(x,t_3,t_5,\ldots)=-2\partial^2_x \log \theta
(\vec V_0(\gamma_1,\ldots,\gamma_g) + x \vec U_1 +
t \vec U_3 + t_5 \vec U_5 + \ldots|b_{ij}) + C(\Gamma),
\eqno (2.10)
$$
where $V_0(\gamma_1,\ldots,\gamma_g)$, $C(\Gamma)$ are some constants,
$b_{ij}=2 \pi i B_{ij}$, $B_{ij}$ is the Riemann matrix (1.4).
\par
General finite-gap solutions are quasiperiodic. The characterization of
periodic solutions in terms of the inverse data is a complicated problem.
\par
A slightly different approach to the inverse problem is more convenient in
some situations (see [28]). Instead of normalizing $\psi(E,0)=1$ for all
$t$ we consider a function $\Psi(\gamma,\vec t)$, $\gamma \in \Gamma$,
$\vec t=(t_1,t_3,t_5,\ldots)$, $t_1=x$, $t_3=t$ such that
\par
1) $L \Psi(\gamma,\vec t) = E(\gamma) \Psi(\gamma,\vec t)$.
\par
2) $\Psi(\gamma,x+\Pi,t_3,t_5,\ldots)=\exp (i \Pi p(\gamma))
\Psi(\gamma,x,t_3,t_5,\ldots)$
\par
3) $\Psi(\gamma,0,0,0,\ldots)=1$
\par
4) $(\partial_{t_{2n+1}} - A_{2n+1}) \Psi(\gamma,\vec t)=0$.
(Symmetries (2.6) is mutually commuting so the condition 4 is self-consistent).
\par
$\Psi(\gamma,\vec t)$ is called Baker-Akhiezer function. It has the following
analytical properties:
\par
Pr.1) $\Psi(\gamma,\vec t)$ is meromorphic in $\Gamma \backslash \infty$.
\par
Pr.2) $\Psi(\gamma,\vec t)$ has simple poles in the points $\gamma_1$,
\ldots, $\gamma_g$ and no other singularities in $\Gamma \backslash \infty$.
\par
Pr.3) $\Psi(\gamma,\vec t)$ has an essential singularity as $\gamma
\rightarrow \infty$
$$
\Psi(\lambda,\vec{t})=\exp(\Theta(\gamma,\vec t))
[1+\chi_1(\vec{t})/\lambda+\chi_2(\vec{t})/\lambda^2+\ldots]
\eqno (2.11)
$$
where in the correspondent to KdV hyperelliptic case $\lambda=\sqrt{-E}$,
$\Theta(\gamma,\vec t)=\lambda x+\lambda^3t+\lambda^5t_5+\ldots$.
\par
{\bf Lemma 2.2.} The properties Pr.1 -Pr.3 uniquely determined the function
$\Psi(\gamma,\vec t)$. It can be expressed in terms of the theta-functions
(see, for example review [33])
$$
\Psi(\gamma,\vec t)=\frac{\theta(\sum t_k \vec U_k + \vec A(\gamma) -
\vec A(\gamma_1) - \ldots - \vec A(\gamma_g) + \vec K)\theta(
- \vec A(\gamma_1) - \ldots - \vec A(\gamma_g) + \vec K)}
{\theta(\sum t_k \vec U_k  -\vec A(\gamma_1) - \ldots - \vec A(\gamma_g) +
\vec K) \theta(\vec A(\gamma) -
\vec A(\gamma_1) - \ldots - \vec A(\gamma_g) + \vec K)} \cdot
$$
$$
\cdot \exp (\sum t_k \int^\gamma \Omega_k).
\eqno (2.12)
$$
Here $\vec K$ is the vector of Riemann constants (see paragraph 1.4),
the integrals are normalized by $\int^\gamma \Omega_k=\lambda^k+o(1)$.
\par
Comparing (2.12) with
$$
u(\vec t) = 2 \partial_x \chi_1(\vec t)
\eqno (2.13)
$$
we can easily derive (2.12).
\par
Now we will recall the finite-gap inverse scattering transform for KP [29].
$$
(u_t - \frac {1}{4} u_{xxx}+\frac{3}{2} uu_x)_x = 3 u_{yy}.
\eqno (2.14)
$$
Let $\Gamma$ be {\em arbitrary} Riemann surface of genus $g$ with a
marked point $\infty$, a local parameter $1/\lambda$ in $\infty$ and
$g$ marked points $\gamma_1$, \ldots, $\gamma_g$. Then for the data of
general position there exists a unique function $\Psi(\gamma,\vec t)$,
$\gamma \in \Gamma$, $\vec t=(t_1,t_2,t_3,t_4,t_5,\ldots)$, $t_1=x$,
$t_2=y$, $t_3=t$ with the analytic properties Pr.1 - Pr.3
(for KP $\Theta(\gamma,\vec t)=\lambda x+\lambda^2 y+\lambda^3 t+
\lambda^4 t_4+\lambda^5 t_5+\ldots$). Representation (2.12) is valid
for general Riemann surfaces but we have to sum in (2.12) by even indexes
as well as by odd ones.
\par
For all $k$ there exists a unique ordinary differential operator $\tilde A_n$
such that
$$
(\partial_{t_n} - \tilde A_n) \Psi(\gamma,\vec t) = 0, \ \
\tilde A_n=\partial_x^{n}+\ldots,\ \tilde A_2=-L.
\eqno (2.15)
$$
The compatibility conditions
$$
[(\partial_{t_2} - \tilde A_2),(\partial_{t_k} - \tilde A_k)]=0
\eqno (2.16)
$$
give the Lax pair for KP as $k=2$ and isospectral symmetries as $k>3$.
The potential $u(\vec t)$ is given by slightly changed (2.10) - the
sum is over all indexes - odd and even. $y$ - independent KP solutions
satisfy KdV.
\par
We need also the Baker-Akhiezer conjugate differential
$\Psi^+(\gamma,\vec t)$. Its analytical properties are the following
\par
Pr.1') $\Psi^+(\gamma,\vec t)$ is a holomorphic in $\Gamma \backslash \infty$
1-differential.
\par
Pr.2') $\Psi^+(\gamma,\vec t)$ has simple zeros in the points $\gamma_1$,
\ldots, $\gamma_g$.
\par
Pr.3') $\Psi^+(\gamma,\vec t)$ has an essential singularity as $\gamma
\rightarrow \infty$
$$
\Psi(\lambda,\vec{t})=d\lambda \ \exp(-\Theta(\gamma,\vec t))
[1+\chi^+_1(\vec{t})/\lambda+\chi^+_2(\vec{t})/\lambda^2+\ldots]
\eqno (2.17)
$$
where $\Theta(\gamma,\vec t)=\lambda x+\lambda^2 y+\lambda^3 t+
\lambda^4 t_4+\lambda^5 t_5+\ldots$.
$$
(\partial_{t_n} + \tilde A^+_n) \Psi^+(\gamma,\vec t) = 0,
\eqno (2.18)
$$
where $\tilde A^+_n$  is the formal conjugate to $\tilde A_n$, i.e.
$(a(x) (\partial_x)^n)^+=(-\partial_x)^n a(x)$. The following ortogonality
properties are important:
\par
{\bf Lemma 2.3.} 1) Let $S$ be a small contour surrounding the point $\infty$.
Then
$$
\oint_S \Psi(\gamma,\vec t) \Psi^+(\gamma,\vec t') = 0
\eqno (2.19)
$$
for all $\vec t$, $\vec t'$.
\par
2) Let $p(\gamma)=\int^\gamma dp$ where $dp$ is the quasimomentum
differential defined in the section 1, $G(\gamma)$ be a contour in $\Gamma$
consisting of all points $\gamma'$ such that $\hbox{Im}\ p(\gamma') =
\hbox{Im}\ p(\gamma)$. Then in the contour $G(\gamma)$ the following relation
is valid
$$
\int_{-\infty}^{+\infty} \Psi(\gamma,\vec t) \Psi^+(\gamma',\vec t) =
2 \pi i \delta(\gamma - \gamma').
\eqno (2.20)
$$
\par
We have to calculate deformations of the Baker - Akhiezer function. For
this purpose we need an appropriate Cauchy kernel.
\par
{\bf Lemma 2.4.} ([20]). The Cauchy - Baker - Akhiezer kernel
$\omega(\gamma,\gamma',\vec t)$ with the following properties:
\par
1) $\omega(\gamma,\gamma',\vec t)$ is a function in $\gamma$ and a 1-form in
$\gamma'$.
\par
2) $\omega(\gamma,\gamma',\vec t)$ is meromorphic function of $\gamma$ in
$\Gamma \backslash \infty$ with simple poles $\gamma_1$, \ldots, $\gamma_g$,
$\gamma'$.
\par
3) As a function of $\gamma'$ $\omega(\gamma,\gamma',\vec t)$ is meromorphic
in $\Gamma \backslash \infty$ with one pole $\gamma$ and zeros in the points
$\gamma_1$, \ldots, $\gamma_g$.
\par
4) $\omega(\gamma,\gamma',\vec t) = o(\exp(\Theta(\gamma,\vec t)))$ as
$\gamma \rightarrow \infty$.
\par
5) $\omega(\gamma,\gamma',\vec t) = o(\exp(-\Theta(\gamma',\vec t))) d \lambda$
as $\gamma' \rightarrow \infty$.
\par
6) $\omega(\gamma,\gamma',\vec t) \sim \frac {d \lambda'}{2 \pi i (\lambda'-
\lambda)}\ \hbox{as} \ \lambda \rightarrow \lambda'$.
\par
\noindent
can be written in the following form
$$
\omega(\gamma,\gamma',x,y,t,\ldots)=\frac{i}{2 \pi} \int_{\pm \infty}^x
\Psi(\gamma,x',y,t,\ldots) \Psi^+(\gamma',x',y,t,\ldots) dx'.
\eqno (2.21)
$$
If $\gamma' \not \in G(\gamma)$ the sing in limit of integration in (2.21)
is uniquely determined by the convergence condition. For
$\gamma' \in G(\gamma)$ the property (2.20) guaranties correctness.
\par
First formula similar to (2.21) was obtained by I.M.Krichever and S.P.Novikov
in [19] for systems with discrete $x$.
\vspace {12pt}
\par
\noindent
{\bf 2.2. Nonisospectral Symmetries. Action On The Finite-Gap Solutions.}
\vspace {12pt}
\par
We have pointed out in the introduction that KdV equation possesses the
following set of symmetries
$$
\frac {\partial u}{\partial \tau_{2n}}=-2\frac {d}{dx}
{\left(-\frac{\Lambda}{4}\right)}
^{n+1}(\sum_{k=0}^{\infty}((2k+1)t_{2k+1}
{\left(-\frac {\Lambda}{4}\right)}^k)\cdot 1.
\eqno (2.22)
$$
\par
They can be written in much more convenient form suggested by A.Yu.Orlov
and E.I.Schulman [17].
$$
\frac {\partial u}{\partial \tau_{2n}}=-2\frac {d}{dx}\ res
\mid_{\lambda =\infty}(\lambda ^{2n+1}\partial _{\lambda}
\Psi(\lambda,\vec t))\Psi^+(\lambda,\vec t),
\eqno (2.23)
$$
\par
Ordinary higher KdV equations (2.4) have similar representation
$$
\frac {\partial u}{\partial _{t_{2n+1}}}=-2\frac {d}{dx}\ res
\mid_{\lambda =\infty}\lambda^{2n+1}\Psi(\lambda,\vec t)\Psi^+(\lambda,\vec t).
\eqno (2.24)
$$
The formulas (2.22), (2.24) can be treated in the following way. Let us
substitute the asymptotical expansions (2.11), (2.17) to the linear problem
$$
L \Psi(\lambda,\vec t) = - \lambda^2 \Psi(\lambda,\vec t),\ \
L \Psi^+(\lambda,\vec t) = - \lambda^2 \Psi^+(\lambda,\vec t).
\eqno (2.25)
$$
 From (2.25) we can calculate all the coefficients $\chi_k(\vec t)$,
$\chi^+_k(\vec t)$ via $u(\vec t)$ (in nonlocal form). Then we substitute
them to (2.23), (2.25). All the exponents in $\Psi \Psi^+$ are reduced
and the residue can be explicitly calculated. We obtain a close system
on $u(\vec t)$ (may be nonlocal).
\par
{\bf Lemma 2.5.} Equations (2.23), (2.25) coincides with (2.22), (2.4)
respectively.
\par
Proof of the Lemma. For small $n$ we can check it by direct calculations.
Then we apply the following identity
$$
\Lambda (\Psi(\lambda,\vec t)\Psi^+(\lambda,\vec t)) = - 4 \lambda^2
(\Psi(\lambda,\vec t)\Psi^+(\lambda,\vec t)).
\eqno (2.26)
$$
(it is a direct consequence of (2.25)).
\par
For the KP equation we can write two-parametric set of symmetries ([17])
$$
\frac {\partial u}{\partial \tau_{nm}}=-2\frac {d}{dx}\ res
\mid_{\lambda =\infty}(\lambda ^{n}\partial^m _{\lambda}
\Psi(\lambda,\vec t))\Psi^+(\lambda,\vec t)
\eqno (2.27)
$$
but only the symmetries with $m=0,1$ are compatible with the finite-gap
structure.
\par
{\bf Theorem 2.1.} Let $\Gamma$ be a Riemann surface with a marked point
$\infty$, a local parameter $1/\lambda$ in $\infty$, a set of points
$\gamma_1$, \ldots, $\gamma_g$ in $\Gamma$ where $g$ is the genus of
$\Gamma$, $l_n=\lambda^{n+1}\partial_\lambda$ be a vector field in the
punctured neighbourhood of $\infty$. Consider the deformation of $\Gamma$
generated by the field $l_n$ (it was described in the paragraph 1.3).
Then the correspondent variation of the KP solution, constructed by this
data reads as
$$
\frac {\partial u}{\partial \tau_{n}}=-2\frac {d}{dx}\ res
\mid_{\lambda =\infty}(\lambda ^{n+1}\partial _{\lambda}
\Psi(\lambda,\vec t))\Psi^+(\lambda,\vec t)
\eqno (2.28)
$$
i.e. it coincides with a symmetry (2.27) such that $m=1$.
\par
Proof of the theorem. In the paragraph 1.3 we have explained that the
calculation of the Baker-Akhiezer function variation is equivalent to the
following Riemann problem on the contour $S$, surrounding the point $\infty$
$$
(\delta \Psi(\lambda,\vec t))_+ -(\delta \Psi(\lambda,\vec t))_- =
\lambda^{n+1}\partial_\lambda \Psi(\lambda,\vec t).
\eqno (2.29)
$$
Solution of (2.29) reads as
$$
\delta \Psi(\lambda,\vec t)= \oint_S
\omega(\lambda,\mu,\vec t) \mu^{n+1} \partial_\mu \Psi(\mu,\vec t).
\eqno (2.30)
$$
where $\omega(\lambda,\mu,\vec t)$ is the Cauchy - Baker - Akhiezer kernel
defined in the Lemma 2.4. Expanding (2.30) as $\lambda \rightarrow \infty$
and using (2.20) we obtain (2.28).
\par
{\bf Corollary 1.} Symmetries (2.23) act on the finite-gap KdV solutions as
$$
\partial E_s/\partial \tau_{2n}=2(-1)^nE_s^{n+1},\
\partial E(\gamma _k)/\partial \tau_{2n}=2(-1)^nE^{n+1}(\gamma _k).
\eqno (2.31)
$$
i.e. all the spectral data is shifted via the vector field
$2(-1)^nE^{n+1}\partial /\partial E$ (see paragraph 1.3).
\vspace{24.0pt}
\par
\noindent
{\bf {3. WHITHAM EQUATIONS. THE FULL ABEL HIERARCHY AND NONISOSPECTRAL
SYMMETRIES.}}
\vspace{12.0pt}
\par
The KdV equation (2.3) and the symmetries (2.6) (they are called higher KdV)
form a commutative set of flows. This set is called KdV hierarchy.
\par
Averaged KdV hierarchy was constructed in [2], [3]. We will not discuss
how the averaged KdV equations can be derived from the original ones and
so we will only recall the answer.
\par
The starting point for the averaging procedure is the space of all $g$-gap
KdV solutions. Such solutions are parameterized by the branch points $E_1$,
\ldots, $E_{2g+1}$ and the points $\gamma_1$, \ldots, $\gamma_g$ in $\Gamma$.
The branch points $E_k$ are integrals of motion and the points $\gamma_k$
play the role of phases (see for example [28] and references therein).
\par
If we consider a slow modulated wave-type solution then the points $E_k$ slowly
depend on coordinate and times $E_k=E_k(X,T,T_5,\ldots)$ where $X=\epsilon x$,
$T=\epsilon t$, $T_5=\epsilon t_5$, \ldots, $\epsilon \ll 1$. Functions
$X$, $T$, $T_5$, \ldots are called slow variables.
\par
For averaging any full set of integrals can be used. But direct calculations
for $g=1$ (see [1]) shows that the variables $E_k$ result in the simplest form
of the averaged equations so it is very natural to use them for higher genera.
\par
The averaged KdV hierarchy can be written in the following form, suggested
by Flaschka - Forest - McLaughlin [2]
$$
\frac{\partial E_k}{\partial T_{2n+1}} = w_k^{2n+1}(E_1,\ldots,E_{2g+1})
\frac{\partial E_k}{\partial X},
\eqno (3.1)
$$
where
$$
w_k^{2n+1}(E_1,\ldots,E_{2g+1}) = \frac {\Omega_{2n+1}(E_k)}{\Omega_1(E_k)},
\eqno (3.2)
$$
$\Omega_k$ are the differentials defined in the paragraph 1.2. We see that
the flows (3.1) have the Riemann diagonal form.
\par
The averaged KdV equations have wider symmetry group then the original KdV.
For example the scaling transform $X \rightarrow \alpha X$,
$T \rightarrow \alpha T$ has no analogs in original equations.
\par
 From the results of S.P.Tsarev [5], [7] it was known that the Whitham
equations
have $2g+1$ infinite series of symmetries but the averaged KdV hierarchy gives
us only one of them. The full set of symmetries was constructed by S.P.Tsarev
in [5] for $g=1$ and B.A.Dubrovin for all $g$. It  contains $2g+1$ infinite
series
plus one finite and can be written as
$$
\frac{\partial E_k}{\partial T^{a_i}_n} = \frac
{\omega^i_n(E_k)}{\Omega_1(E_k)}
\frac{\partial E_k}{\partial X},
$$
$$
\frac{\partial E_k}{\partial T^{b_i}_n} = \frac
{\sigma^i_n(E_k)}{\Omega_1(E_k)}
\frac{\partial E_k}{\partial X},
\eqno (3.3)
$$
$$
\frac{\partial E_k}{\partial T^H_i} = \frac {\omega_i(E_k)}{\Omega_1(E_k)}
\frac{\partial E_k}{\partial X},
$$
where $i=1,\ldots,g$, $n=1,2,\ldots$. Let us denote
$$
w_k^{a_i n} = \frac {\omega^{a_i}_n(E_k)}{\Omega_1(E_k)},
w_k^{b_i n} = \frac {\omega^{b_i}_n(E_k)}{\Omega_1(E_k)},
w_k^{H n} = \frac {\omega_i(E_k)}{\Omega_1(E_k)}.
\eqno (3.4)
$$
(see the definitions in the paragraph 1.2).
\par
We have four families of differentials, times and velocities. To avoid too long
notations we will use the following agreement:
\par
$\Omega^F_\alpha$ may denote {\em any} of the differentials $\Omega_k$,
$\omega_i$, $\omega^i_k$, $\sigma^i_k$;\ $w^{F \alpha}_s$ and $T^F_\alpha$
are the correspondent velocities and times, $Q^F_{\alpha \beta} =
V_{\Omega^F_\alpha \Omega^F_\beta}$ where $V$ is the scalar product defined in
the paragraph 1.2.
\par
The following statements are important for us.
\par
{\bf Lemma 3.1} ([2], [6]). The flows (3.1), (3.3) can be written as
$$
\frac{\partial \Omega_1}{\partial T^F_\alpha}=
\frac{\partial \Omega^F_\alpha}{\partial X}.
\eqno (3.5)
$$
Here we always assume
$$
\frac{\partial}{\partial T^F_\alpha}=
\frac{\partial}{\partial T^F_\alpha}\mid_{E=\hbox{const}}.
$$
\par
Proof of the Lemma. The differentials in the both sides of (3.5) have the
following properties.
\par
1) They are single-valued in $\Gamma$.
\par
2) Their integrals by $a_j$-cycles are equal to 0.
\par
3) They have second-order poles in the branch points $E_1$, \ldots, $E_{2g+1}$
and no other singularities. All their residues are equal to zero.
\par
4) Let $z=\sqrt{E-E_k}$ be local parameter in the neighbourhood of the branch
point $E_k$. Then the singular part of both differentials is equal to
$\Omega^F_\alpha(z)/2z^2$.
\par
Properties 1-4 uniquely determined a differential so the left-hand side is
equal to the right-hand one.
\par
{\bf Lemma 3.2} ([2], [6]). All the symmetries (3.1), (3.3) are mutually
commuting.
\par
Proof of the Lemma. Consider the differentials
$\partial \Omega^F_\alpha/\partial T^F_\beta$,
$\partial \Omega^F_\beta/\partial T^F_\alpha$.
They satisfy the properties 1-3 of the Lemma 3.1 and have the same
singularities in the branch point so they are equal
$$
\frac{\partial \Omega^F_\alpha}{\partial T^F_\beta}=
\frac{\partial \Omega^F_\beta}{\partial T^F_\alpha}.
$$
and we have
$$
\frac{\partial} {\partial T^F_\alpha} \frac{\partial} {\partial T^F_\beta}
\Omega^F_1=
\frac{\partial}{\partial X}
\frac{\partial \Omega^F_\beta}{\partial T^F_\alpha}=
\frac{\partial}{\partial X}
\frac{\partial \Omega^F_\alpha}{\partial T^F_\beta}=
\frac{\partial} {\partial T^F_\beta} \frac{\partial} {\partial T^F_\alpha}
\Omega^F_1.
$$
\par
{\bf Lemma 3.3} ([8], [23], [6]). The scalar products $Q^F_{\alpha\beta}$
satisfy the following relation
$$
\frac{\partial Q^F_{\alpha\beta}}{\partial T^F_\gamma}=
\frac{\partial Q^F_{\alpha\gamma}}{\partial T^F_\beta}.
\eqno (3.6)
$$
Thus we can define functions
$$
Q^F_\alpha(\vec T^F)= \int_{\vec 0}^{\vec T^F} \sum_\beta Q^F_{\alpha\beta}
d T^F_\beta.
\eqno (3.7)
$$
Here the vector $\vec T^F$ contains all the times.
\par
Proof of the Lemma. From the formulas (1.21) we see that the coefficients
$\partial Q^F_{\alpha\beta}/\partial T^F_\gamma$ can be expressed as
integrals or expansion coefficients for the differential
$\partial \Omega^F_{\beta}/\partial T^F_\gamma$ Applying the Lemma
3.2 we complete the proof.
\par
{\bf Lemma 3.4} Let us shift the branch points $E_s$ of the surface
$\Gamma$ and the normalization point $E_0$ via a vector field
$l_{2k}=2 (-1)^k E_{k+1} \partial_E$
$$
\frac{\partial E_s}{\partial \tau_{2n}} = 2 (-1)^n E_s^{n+1},\ \
s=0,1,\ldots,2g+1.
\eqno (3.8)
$$
Then we have
$$
\frac{\partial w_s^{2k+1}}{\partial \tau_{2n}} =
(2k+1)w_s^{2k+2n+1} - w_s^{2n+1} w_s^{2k+1} + \sum _{m=0}^{n-1} \left [ \left(
\frac{\partial}{\partial T_{2k+1}} - w_s^{2k+1}\frac{\partial}{\partial X}
\right ) Q_{2m+1} \right ] w_s^{2n-2m-1}.
\eqno (3.9)
$$
$$
\frac{\partial w_s^{Hk}}{\partial \tau_{2n}} =
- w_s^{2n+1} w_s^{Hk} + \sum _{m=0}^{n-1} \left [ \left(
\frac{\partial}{\partial T^H_{k}} - w_s^{Hk}\frac{\partial}{\partial X}
\right ) Q_{2m+1} \right ] w_s^{2n-2m-1}.
\eqno (3.10)
$$
$$
\frac{\partial w_s^{a_ik}}{\partial \tau_{2n}} =
2 (-1)^n kw_s^{a_ik+n} - w_s^{2n+1} w_s^{a_ik} + \sum _{m=0}^{n-1}\left[\left(
\frac{\partial}{\partial T^{a_i}_{k}} - w_s^{a_ik}\frac{\partial}{\partial X}
\right ) Q_{2m+1} \right ] w_s^{2n-2m-1}.
\eqno (3.11)
$$
$$
\frac{\partial w_s^{b_ik}}{\partial \tau_{2n}} =
2 (-1)^n kw_s^{b_ik+n} - w_s^{2n+1} w_s^{b_ik} + \sum _{m=0}^{n-1}\left[\left(
\frac{\partial}{\partial T^{b_i}_{k}} - w_s^{b_ik}\frac{\partial}{\partial X}
\right ) Q_{2m+1} \right ] w_s^{2n-2m-1}.
\eqno (3.12)
$$
(The functions $Q_j$ were defined in Lemma 3.3).
\par
The Lemma is proved by direct calculations using (1.35) - (1.38) and (3.2),
(3.4).
\par
{\bf Lemma 3.5.} Let us have a pair of flows
$$
\frac{\partial E_s}{\partial \tau} = R_s(E_s) + \left ( \sum_\alpha
f_\alpha(\vec T) w_s^{F\alpha} \right ) \frac{\partial E_s}{\partial X},
\eqno (3.13)
$$
$$
\frac{\partial E_s}{\partial T^F_\gamma} =  w_s^{F\gamma}
\frac{\partial E_s}{\partial X}.
\eqno (3.14)
$$
\par
Then the flows (3.13) and (3.14) commute if and only if the following
compatibility condition holds:
$$
\frac{\partial w_s^{F\gamma}}{\partial \tau} =
\sum_\alpha \left [ \left (
\frac{\partial}{\partial T^F_{\gamma}} - w_s^{F\gamma}
\frac{\partial}{\partial X} \right ) f_\alpha(\vec T) \right ]
w_s^{F\alpha}
\eqno (3.15)
$$
where $\partial w_s^{F\gamma}/\partial \tau$ denotes the variations of
velocities via the shift
$$
\frac{\partial E_s}{\partial \tau} = R_s(E_s).
\eqno (3.16)
$$
\par
This Lemma is proved by simple direct calculation.
\par
Now we have prepared everything to formulate and prove our main result.
\par
{\bf Theorem 3.1.} The flows
$$
\frac{\partial E_s}{\partial \tau_{2n}} = 2(-1)^nE_s^{n+1}+
\left \{ \sum_{k\ge 0} (2k+1)T_{2k+1}w_s^{2k+2n+1} \right \}
\frac {\partial E_s}{\partial X}+
$$
$$
+ \left\{ 2 (-1)^n \sum_{k\ge 0}
 k (T_k^{a_i} w_s^{a_ik+n}+T_k^{b_i} w_s^{b_ik+n})
+ \sum_{0\le k<n}Q_{2k+1}w_s^{2n-2k-1}  \right \}
\cdot \frac {\partial E_s}{\partial X}.
\eqno (3.17)
$$
commute with the whole Whitham hierarchy (3.1), (3.3). The nonlocal functions
$Q_j$ were defined in the Lemma 3.3. All the partial derivatives of $Q_j$
can be expressed via branch points $E_s$ and normalization point $E_0$ in terms
of hyperelliptic integrals (see Lemma 1.1).
\par
Comparing the Lemmas 3.4 and 3.5 we prove this theorem.
\vspace{12pt}
\par
The author is very grateful to B.A.Dubrovin, I.M.Krichever and S.P.Tsarev
for useful discussions.
\vspace{24pt}
\par
\noindent
{\bf REFERENCES.}
\vspace{12pt}
\par
1. G.B.Whitham. Linear and nonlinear waves. N.Y., Wiley, (1974).
\par
2. H.Flaschka, G.Forest, D.W.McLaughlin. Commun. Pure Appl. Math.
{\underline {33}}, no.6 (1980).
\par
3. Peter D.Lax, C.David Levermore. Commun. Pure Appl. Math.,
{\underline {36}}, pp.253-290, 571-593, 809-830 (1983).
\par
4. B.A.Dubrovin, S.P.Novikov. Sov. Math. Doklady {\underline{27}},
p.665 (1983); Russian Math. Surveys 44:6  (1989), p.35.
\par
5. S.P.Tsarev. Math. USSR Izvestiya, {\underline {54}}, no. 5 (1990).
\par
6. B.A.Dubrovin. Hamiltonian  formalism of Whitham-type hierarchies
and topological Landau-Ginzburg models. Preprint 1991.
\par
7. S.P.Tsarev. Soviet. Math. Dokl. {\underline {31}}, pp.488-491 (1985).
\par
8. I.M.Krichever. Funct. Anal. Appl. {\underline{11}} (1988), p.15.
\par
9. F.Calogero, A.Degasperis. Lett. Nuov. Cim. {\underline {22}}, p.420 (1978).
\par
10. V.A.Belinskii, V.E.Zakharov. JETP {\underline {12}}, p.1953 (1978).
\par
11. Maison. Phys. Rev. Lett. {\underline {41}}, p.521 (1978).
\par
12. G.Baricci, T.Regge J. Math. Phys. {\underline {18}} No 6, p.1149 (1976).
\par
13. N.H.Ibragimov, A.B.Shabat. Dokl. Akad. Nauk SSSR {\underline {244}} (1979),
p.1.
\par
14. H.H.Chen, Y.C.Lee, J.E.Lin. Physica 9D, {\underline 9}, No 3, pp.439-445
(1983).
\par
15. F.J.Schwarz. J. Phys. Soc. Japan, {\underline {51}} No 8, p. 2387 (1982).
\par
16. B.Fuchssteiner. Progr. Thoer. Phys., {\underline {70}}, pp. 1508-1522
(1983); W.Oevel, B.Fuchssteiner Phys. Lett. {\underline {88}}A, p.323 (1982).
\par
17. A.Yu.Orlov, E.I.Schulman. Additional Symmetries of the Integrable
Systems and Conformal Algebra Representations. Preprint IA and E, No 217,
Novosibirsk (1984); A.Yu.Orlov, E.I.Schulman. Theor. Math. Phys.,
{\underline {64}}, pp. 323-327 (1985) (Russian); A.Yu.Orlov, E.I.Schulman.
Additional Symmetries for 2-D Integrable Systems. Preprint IA and E, No 277,
Novosibirsk (1985); A.Yu.Orlov, E.I.Schulman. Lett. Math. Phys. {\underline
{12}}, pp. 171-179 (1986). A.Yu.Orlov in Proceedings Int. Workshop "Plasma
theory and \ldots " in Kiev 1987, World Scientific (1988) p.116-134.
\par
18. A.S.Fokas, P.M.Santini. Commun. Math. Phys. {\underline {116}},
p.449 (1088).
\par
19. I.M.Krichever, S.P.Novikov. Funct. Anal. i Prilog. {\underline {21}},
No 2, pp. 46-63 (1987); {\underline {21}}, No 4, pp. 47-61 (1987);
{\underline {23}}, No 1, pp. 41-56 (1989).
\par
20. P.G.Grinevich, A.Yu.Orlov. Virasoro action on Riemann surfaces,
Grassmannians, $\det \bar{\partial_j}$ and Segal-Wilson $\tau$ function.
in "Problmes of modern quantum field theory", ed. A.A.Belavin, A.U.Klimyk,
A.B.Zamolodchikov, Springer 1989.
\par
21. P.G.Grinevich, A.Yu.Orlov. Funct. Anal. Appl. {\underline{24}}
no. 1, (1990); Higher symmetries of KP equation and the Virasoro
action on Riemann surfaces. in "Nonlinear evolution equations and
dynamical systems", ed. S.Carillo, O.Ragnisco, Springer 1990.
\par
22. A.Gerasimov, A.Marshakov, A.Mironov, A.Morozov, A.Orlov. Nuclear Phys.
B {\underline {357}}, pp. 565-618 (1991).
\par
23. I.M.Krichever. "The dispersionless Lax equations and topological
minimal models". Preprint 1991 Torino; Lecture at the Sakharov memory
congress (1991).
\par
24. G.Springer. Introduction to Riemann surfaces. Addison - Wesley
Publishing Company, Inc. Reading, Massachusetts, USA (1957).
\par
25. M.Shiffer, D.K.Spencer. Functionals of finite Riemann surfaces.
Princeton, New Jersey (1953).
\par
26. D.Mumford. Tata lectures on Theta. Birkh\"auser, Boston - Basel -
Stuttgart (1983).
\par
27. R.Dijkgraaf, H.Verlinde, E.Verlinde. "Topological strings in $d<1$".
Princeton preprint PUPT - 1024, IASSNS-HEP - 90/71.
\par
28. V.E.Zakharov, S.V. Manakov, S.P. Novikov, L.P. Pitaevsky. Soliton theory.
Plenum, New York, 1984.
\par
29. I.M.Krichever. Docl. Akad. Nauk. SSSR, {\underline {227}}, No 2, pp.
291-294 (1977).
\par
30. I.M.Krichever. Uspekhy. Mat. Nauk. {\underline {44}}, No 2, pp. 121-184
(1989).
\par
31. I.M.Gel'fand, L.A.Dikii. Funct. Anal. Pril. {\underline {10}}, No 4,
pp. 13-29 (1976).
\par
32. H.McKean, E.Trubowitz. Commun. Pure. Appl. Math. {\underline{29}},
pp. 143-226 (1976).
\par
33. B.A.Dubrovin. Uspekhi. Math. Nauk. {\underline {36}}, No 2, pp. 11-40
(1981).
\par
34. N.Kawamoto, Yu.Namikava, A.Tsuchiya, Ya.Yamada. Commun. Math. Phys.
{\underline {116}}, No. 2, pp.247-308 (1988).
\end{document}